\documentstyle[12pt,titlepage]{article} 

\title{A Theoretical Prediction for Exclusive Decays $B \rightarrow
K(K^*)
\eta_c}
\author{Mohammad R. Ahmady and Roberto R. Mendel}
\date{December, 1993}   
\def\_{\rule{.3em}{.15ex}}  

\begin{document}           
\begin{titlepage}
 \begin{center}
  \vspace{0.75in}
  {\bf {\LARGE A Theoretical Prediction for Exclusive Decays
  $B \rightarrow K(K^*) \eta_c$} \\
  \vspace{0.75in}
  Mohammad R. Ahmady and Roberto R. Mendel} \\
  Department of Applied Mathematics\\
  The University of Western Ontario \\
  London, Ontario, Canada\\
  \vspace{1in}
  ABSTRACT \\
  \vspace{0.5in}
  \end{center}
  \begin{quotation}
  \noindent The decay rates for the exclusive B decays $B \rightarrow
K \eta_c
  $ and $B \rightarrow K^* \eta_c$ are calculated in the context of
the  heavy quark
  effective theory.  We obtain $\Gamma (B \rightarrow K \eta_c
)/\Gamma (B
\rightarrow K\psi )=1.6 \pm 0.2$ and $\Gamma (B\rightarrow K^*\eta_c
)/\Gamma (B \rightarrow K^*\psi )=0.39 \pm 0.04$. These results lead
to estimates
$BR(B \rightarrow K
\eta_c)=(0.11 \pm 0.02) \%
    $ and $BR(B \rightarrow K^* \eta_c)= (0.05 \pm 0.01) \% $ if we
use the
central
current experimental values for $B\rightarrow (K,K^*)\psi $ branching
ratios.

  \end{quotation}
\end{titlepage}


\newcommand{\da}{\mbox{$\scriptscriptstyle \dag$}}
\newcommand{\lag}{\mbox{$\cal L$}}
\newcommand{\tr}{\mbox{\rm Tr\space}}
\newcommand{\fc}{\mbox{${\widetilde F}_\pi ^2$}}
\newcommand{\ns}{\textstyle}
\newcommand{\si}{\scriptstyle}
With the recent progress in the measurement of some nonleptonic B
decay
channels\cite{CLEO}, the exclusive decays $B \rightarrow K (K^*)
\eta_c$ could be
within experimental reach very soon.  In this paper, we calculate the
decay
rate of the above processes in the context of the heavy quark
effective theory.
The branching ratio for these channels are obtained in terms of
decay rates $\Gamma (B \rightarrow K(K^*) \psi
)$, which have been measured \cite {CLEO}

The effective Hamiltonian relevant to the process $ b
\rightarrow s c \bar c$ can be written as \cite{DTP}:
\begin{equation}
 H _{eff} = C  \bar s \gamma _{\mu} (1- \gamma _5 )b
 \bar c \gamma _{\mu} (1- \gamma _5 )c \;\; ,
\end{equation}
where $$
 C= {{G _F} \over {\sqrt 2}} (c _1 + c _2 /3 ) V _{cs} ^* V _{cb}
\;\; .
$$
 $c _1 $ and $c _2$ are QCD improved coefficients, $V_{cs} {\rm and}
V_{cb}$
 are elements of CKM mixing matrix.
 Assuming factorization and using the definition for
$f_{\eta_c}$, the
 decay constant for the psuedoscalar charmonium state:
 $$ f_{\eta_c} q ^{\mu}  = <0| \bar c \gamma ^{\mu} \gamma_5
 c| \eta_c > \;\; , $$ \\
 leads to $H_{eff}$ for $B \rightarrow X_s \eta_c$, where $X_s$ is
 hadronic final state containing a strange quark
 $$  H _{eff} = C f_{\eta_c} \bar s \gamma _{\mu} (1- \gamma _5 )b
q^\mu \;\; .
 $$
To calculate the exclusive decay rates $ \Gamma (B \rightarrow
K \eta_c ) $ and $ \Gamma (B \rightarrow K^* \eta_c )$ we need
to evaluate the following matrix elements:
\begin{equation}
<K(p^{\prime} )| \bar s \gamma _{\mu} (1- \gamma _5 )b |B(p)>\;\; ,
\end{equation}
\begin{equation}
<K^* (p^{\prime} , \epsilon )| \bar s \gamma _{\mu} (1- \gamma _5 )b
|B(p)>
\;\; .
\end{equation}
Using heavy quark effective theory, i.e. assuming that bottom and
strange
quark are both heavy compare to $ \Lambda _{QCD} $, we can write (2)
and
(3) in terms of one universal Isgur-Wise function \cite{IW}:
\begin{equation}
<K(v^{\prime} )| \bar s \gamma _{\mu} (1- \gamma _5 )b |B(v)> = \sqrt
{m _K m_ B } \xi (v. v^{\prime} ) ( v _{\mu} + v _{\mu} ^{\prime}
)\;\; ,
\end{equation}
\begin{equation}
\begin{array}{l}
<K^* (v^{\prime} , \epsilon )| \bar s \gamma _{\mu} (1- \gamma _5 )b
|B(v)> = \\ \sqrt {m _{K^*} m_ B } \xi (v. v^{\prime} ) \left [ (1+v.
v^{\prime} ) \epsilon _{\mu} - v _{\mu} ^{\prime} ( \epsilon . v ) +
i
\epsilon _{\mu \alpha \beta \gamma } v^{\alpha} v^{\prime \beta}
\epsilon ^{\gamma} \right ]\;\; .
\end{array}
\end{equation}
where $ v $ and $ v ^{\prime}$ are velocities of initial and final
state mesons respectively. Consequently, the decay rates are as
follows:
\begin{equation}
\Gamma (B \rightarrow K \eta_c ) = \frac {C^2 f^2 _{\eta_c}}{16 \pi }
g(m _B , m _K , m _{\eta_c} )  m_K m_B ^2 {\left (1- \frac {m_K}{m_B}
\right
)}^2 {\left (1+v. v^\prime \right )}^2 {\vert \xi (v. v ^{\prime} )
\vert }^2\;\; ,
\end{equation}

\begin{equation}
\Gamma (B \rightarrow K^* \eta_c ) = \frac {C^2 f^2 _{\eta_c}}{16 \pi
}
g(m _B , m _{K^*} , m _{\eta_c} )  m_{K^*} m_B ^2 {\left (1+ \frac
{m_{K^*}}{m_B} \right
)}^2 \left ({v. v^\prime }^2 -1 \right ) {\vert \xi (v. v ^{\prime} )
\vert }^2 \; ,
\end{equation}
where
$$ g(a , b , c ) =  {\left [ {\left (1- \frac
{c^2 }{a^2} -\frac {b^2}{a^2} \right )}^2 - \frac
{4 b^2 c^2 }{a^4} \right ]}^{1/2} \; ,$$
and
$$v. v ^{\prime} = \frac {m^2 _B + m^2 _{K^{(*)}} - m^2 _{\eta_c} }{2
m _B
m _{K^{(*)}} }\;\; . $$
The value of $v.v^\prime $ for $B \rightarrow K \eta_c$ and $B
\rightarrow
 K^* \eta_c$ is 3.69 and 2.04 respectively.

The decay rates for similar processes involving $\psi$ has been
calculated in
reference \cite{AL}
\begin{equation}
\begin{array}{l}
\Gamma (B \rightarrow K \psi ) = \displaystyle{\frac {C^2 f^2 _{\psi}}
{16 \pi } g(m _B , m _K , m _{\psi} )   m _K} \times \\
\displaystyle{\left [{(1+v.v^\prime )}^2 \frac {{(m_B-m_K)}^2}{m_\psi
^2} -2
(1+v.v^\prime ) \right ] {\vert \xi (v.v^\prime ) \vert }^2} \;\; ,\\
\end{array}
\end{equation}
\begin{equation}
\begin{array}{l}
\Gamma (B \rightarrow K^* \psi ) = \displaystyle{\frac {C^2 f^2
_{\psi}}
{16 \pi } g(m _B , m _{K^*} , m _{\psi} )   m _{K^*}} \times \\
\displaystyle{\left [({v.v^\prime }^2 -1) \frac
{{(m_B+m_{K^*})}^2}{m_\psi ^2} +
2(1+2v.v^\prime )(1+v.v^\prime ) \right ] {\vert \xi (v.v^\prime )
\vert
}^2}\;\; ,\\
\end{array}
\end{equation}
where
$$v. v ^{\prime} = \frac {m^2 _B + m^2 _{K^{(*)}} - m^2 _{\psi} }{2 m
_B
m _{K^{(*)}} }\;\; . $$
$v.v^\prime \approx 3.55$ for $B\rightarrow K \psi$ and $\approx 2.02
$ for
$B \rightarrow K^* \psi$.  Using (6), (7), (8) and (9) we obtain:
\begin{equation}
\frac {\Gamma (B \rightarrow K \eta_c)}{\Gamma (B \rightarrow K \psi
)} =
(13.0 {\rm GeV^2}) \frac {{\vert \xi (v.v^\prime =3.69 ) \vert
}^2}{{\vert \xi
(v.v^\prime =3.55)
\vert }^2} \frac {f^2_{\eta_c}}{f^2_{\psi}}\;\; ,
\end{equation}
\begin{equation}
\frac {\Gamma (B \rightarrow K^* \eta_c)}{\Gamma (B \rightarrow K^*
\psi )} =
(3.2 {\rm GeV^2}) \frac {{\vert \xi (v.v^\prime =2.10) \vert
}^2}{{\vert \xi
(v.v^\prime =2.02)
\vert }^2} \frac {f^2_{\eta_c}}{f^2_{\psi}}\;\; .
\end{equation}
We would like to point out that although the strange quark is not
particularly heavy and significant corrections to eqns. (4) and (5)
are expected, these corrections are likely to cancel
out in the ratios (10) and (11).

The ratio of the Isgur-Wise functions in (10) and (11) are expected
to be close
to one as their arguments are nearly equal.  For example, taking
\begin{equation}
\xi (v. v^\prime ) = exp \left [ \frac {9}{256 {\beta}^2} m^2
_{K^{(*)}}
(1-{(v.v^\prime )}^2) \right ]\;\; ,
\end{equation}
which is obtained using the model of Isgur, Wise, Grinstein and Scora
\cite{IWGS}
 and the parameter $\beta = 0.295$ GeV is fixed by the best fit to
the
 measured decay rates $B \rightarrow (K, K^*) +( \psi , \psi^\prime )$
\cite{AL2}
 , this ratio is evaluated to be:
\begin{equation}
\begin{array}{l}
\displaystyle {\frac {{\vert \xi (v.v^\prime =3.69 ) \vert
}^2}{{\vert \xi
(v.v^\prime =3.55) \vert }^2} \approx 0.82\;\; ,} \\
\displaystyle {\frac {{\vert \xi (v.v^\prime =2.10) \vert }^2}{{\vert
\xi
(v.v^\prime =2.02) \vert }^2} \approx 0.81\;\; ,}
\end{array}
\end{equation}
 leading to
\begin{equation}
\frac {\Gamma (B \rightarrow K \eta_c)}{\Gamma (B \rightarrow K \psi
)} =
(10.7 {\rm GeV^2})\frac {f^2_{\eta_c}}{f^2_{\psi}}\;\; ,
\end{equation}
\begin{equation}
\frac {\Gamma (B \rightarrow K^* \eta_c)}{\Gamma (B \rightarrow K^*
\psi )} =
(2.6 {\rm GeV^2})\frac {f^2_{\eta_c}}{f^2_{\psi}}\;\; .
\end{equation}
 The numerical coefficient in (14) and (15) differs from those
obtained from
 pole approximation and BSW model \cite {DT}.
 To evaluate the ratio of the decay constants, we use the potantial
model
 relations relating these constants to the value of the meson
wavefunction
 at the origin:
 \begin{equation}
 \begin{array}{l}
 f_{\eta_c}= \displaystyle{\sqrt {\frac {12}{m_{\eta_c}}}
\Psi_{\eta_c}
(0)}\;\; , \\
 f_\psi = \displaystyle{\sqrt {12 m_\psi } \Psi_{\psi} (0)}\;\; .
 \end{array}
 \end{equation}
The wavefunctions of pseudoscalar meson $\eta_c$ and vector meson
$\psi$ are
not expected to be very different.  Using a simple perturbation
theory argument, we make an estimate \cite{AM}

\begin{equation}
\frac {{\vert \Psi _{\eta_c}(0) \vert}^2}{{\vert\Psi_\psi
(0)\vert}^2}=1.4 \pm 0.1 \;\; .
\end{equation}

Consequently, the ratio of the decay constants turns out to be:
\begin{equation}
\frac {f^2 _{\eta_c}}{f^2 _{\psi}} = \frac {1}{m_{\eta_c} m_\psi }
\frac {{\vert \Psi _{\eta_c}(0) \vert}^2}{{\vert\Psi_\psi
(0)\vert}^2}
 \approx
0.15 \pm 0.01 {\rm GeV^{-2}}\;\; .
\end{equation}
Inserting (13) and (15) in (14) and (15) we obtain:
\begin{equation}
\begin{array}{l}
\displaystyle {\frac {\Gamma (B\rightarrow K\eta_c )}{\Gamma
(B\rightarrow K\psi )}=1.6 \pm 0.2\;\; ,} \\
\displaystyle {\frac {\Gamma (B\rightarrow K^*\eta_c )}{\Gamma
(B\rightarrow K^*\psi )}=0.39 \pm 0.04\;\; .}
\end{array}
\end{equation}
 Using the central values of the measurements \cite{PDG, COM}
$$ BR (B \rightarrow K \psi ) = 0.071 \% \;\; ,$$
and $$
BR (B \rightarrow K^* \psi ) =0.135 \% \;\; ,$$
result in our predictions:
\begin{equation}
\begin{array}{l}
BR (B \rightarrow K \eta_c) = (0.11 \pm 0.02) \% \;\; ,\\
BR (B \rightarrow K^* \eta_c)= (0.05 \pm 0.01) \% \;\; .
\end{array}
\end{equation}

In conclusion, we estimated the ratios of the two body nonleptonic
decay rates of the B meson to $\eta_c$ and $\psi$ for $K$ and $K^*$
final states using heavy quark effective theory.  For $K$ final
state, the decay rate to $\eta_c$ is larger than the
decay to $\psi$ by a factor $\approx 1.6$.  However, for $K^*$ final
state, it is $\psi$ that
is produced more than twice as often as $\eta_c$.  Finally, due to
likely
cancellations in the estimated ratios, we do not expect significant
corrections to our results because of our neglect of subleading terms
in the
heavy quark expansion or our use of a non-relativistic formalism to
relate $f_{\eta_c}$ to $f_\psi$.
\\
\\
{\bf \Large Acknowledgements}\\
This work was supported in part by the Natural Sciences and
Engineering Research Council of Canada.  M. A. would like to thank
the Department of Applied Mathematics at UWO for warm hospitality
during his visit.


\begin{thebibliography}{99}
\bibitem{COM} We have used the charge averages $\Gamma (B\rightarrow
K(K^*)\psi )=1/2 [\Gamma (B^+\rightarrow K^+(K^{*+})\psi
)+\Gamma (B^\circ \rightarrow K^\circ (K^{*\circ})\psi ) ]$ from
Ref. [9].
\end{thebibliography}
\end{document}